\newcommand{\ignore}[1]{}
\ignore{
}

\documentclass[11pt]{article}
\usepackage[xdvi]{epsfig}
\usepackage{amssymb,amsmath}

\newcommand{\half}{ {\scriptstyle \frac{1}{2} } }
\newcommand{\quarter}{ {\scriptstyle \frac{1}{4} } }
\newcommand\be{\begin{equation}}
\newcommand\ee{\end{equation}}
\newcommand\bea{\begin{eqnarray}}
\newcommand\eea{\end{eqnarray}}

\title{Probing the 5th Dimension with  the QCD String
\thanks{Brown-HET-1431:
This work was supported in part by the Department of Energy under
Contracts No. DE-FG02-91ER40676 and No. DE-FG02-91ER40688}}

\author{Richard  C. Brower
\\ Physics Department\\Boston University\\
Boston, MA 02215, USA \\
\and
Chung-I Tan \\
Physics Department \\
Brown University\\
Providence, RI 02912, USA\\
\and 
Ethan Thompson \\
Department of Physics\\
University of Washington \\
Seattle, WA 98195, USA \\
}

\begin{document}

\maketitle

\begin{abstract}
  
  A salient feature of String/Gauge duality is an extra 5th dimension.  Here
  we study the effect of confining deformations of $AdS^5$ and compute the
  spectrum of a string stretched between infinitely massive quarks and
  compare it with the quantum states of the QCD flux as determined by Kuti,
  Juge and Morningstar in lattice simulations.  In the long flux tube limit
  the AdS string probes the metric near the IR cutoff of the 5th dimension with a
  spectrum approximated by a Nambu-Goto string in 4-d flat space, whereas at
  short distance the string moves to the UV region with a  discrete spectrum for
  pure $AdS^5$.  We also review earlier results on glueballs states and the
  cross-over between hard and soft diffractive scattering that support this
  picture.
\end{abstract}

\newpage

\section{Introduction}

The picture for QCD based on String/Gauge duality (see below for some
technical details) proceeds as follows. The flux tube of QCD is dual to a
fundamental string (or world sheet sigma model) in 5 dimensional AdS like
space plus 5 compact dimensions for the critical superstring.  The motion of
the string in 4-d space time $x^\mu$ is accompanied by motion in a transverse
``radial'' co-ordinate, $r$.  The inverse radial co-ordinate $z = R^2/r$
(scaled by the AdS radius $R$) represents conformal transformations in the
Yang-Mills theory. UV physics is mapped into small z (large r) and IR physics
into large z (small r).  The minimal energy of the stretched string is
determined by the geodesic in the x-z space. Its quantum fluctuations give
the spectrum for the flux tube of Yang Mills theory.

Remarkably we find that these transverse fluctuations obey a simple wave equation,
\be
[\; \partial^2_t - v^2(\sigma) \partial^2_\sigma\; ]\; X^\perp = 0 \; ,
\ee
where $\sigma$ is the proper distance along the flux tube as seen by the Yang-Mills 
observer. Independent of the precise warping of the space in the 5th
co-ordinate, the local velocity takes a universal form,
$$v(\sigma) = \frac{z^2(\sigma)}{z^2_c} \; .$$
It is given by the square of the displacement in $z(\sigma)$ of the minimal
surface in the $x\mbox{-}z$ plane. For a long flux tube relative to the
confinement scale, the velocity is nearly constant and equal to the speed of
light ($z \simeq z_c$) so the physics is well approximated by a naive
massless (Nambu-Goto) string in 4-d with energy levels $\Delta E_n \simeq
\pi( n - 1/12)/L + O(1/L^3)$. However, the signal slows as it enters the
environment of the massive quark.  For small $Q\mbox{-}\bar Q$ separation,
the classical energy is Coulombic and the fluctuations are controlled by the
CFT limit.

The goal of this talk is to show how this picture transpires.

\section{String/Gauge Duality}

In 1998 Maldacena~\cite{maldacena,witten,gkp} took a decisive step toward
realizing the ``ancient'' quest for the QCD string.  Under specific
circumstances he gave strong arguments for the ``String/Gauge'' conjecture of
an exact equivalence (aka duality) between 10-d superstring theory in a curved
background and super Yang-Mills theory in 4-d.  Subsequent computations in
the last few years have given hundreds of specific tests of this conjecture
and a myriad of extensions to a large class of Yang-Mills theories.  This
cast the question of the existence of the QCD string in an entirely
new light. We now have a qualitative road map for how such an equivalence may
be realized without to date a precise construction of the string dual to QCD.
It is interesting to explore the physical consequences of this qualitative picture for
QCD, trying to extract features that are independent of the detailed
realization not yet available for analysis.

Maldacena's original celebrated example of the duality between IIB string
theory in an $AdS^5 \times S^5$ background metric,
\be \label{eq:adsmetric}
ds^2 = \frac{r^2}{R^2} \eta_{\mu\nu} dX^\mu dX^\nu + \frac{R^2}{r^2}(
dr^2 + r^2 d^2\Omega_5)\; , 
\ee
and ${\cal N} = 4$ Super Yang-Mills still provides a useful
prototype. The essential feature is the addition of a ``5th radial''
dimension, $r$, transverse to the 3+1 Minkowski space, $X^\mu$.  Large r is
UV physics for the dual Yang-Mills, while small r is the IR region.  In
$AdS^5 \times S^5$ the isometries of the background, $SO(3,2) \times SO(6)$,  
are dual to the conformal  group times the  R symmetries of ${\cal N} =4$ SUSY
Yang-Mills.  The  'tHooft coupling,
\be
\lambda \equiv g^2 N_c = 4\pi g_s N=  R^4/l^4_s \; ,
\ee
in Yang-Mills theory is expressed in terms of the closed string
coupling  $g_s$, the AdS radius $R$ and the string length $l_s$.  (The open
string tension is $T_s = ( 2 \pi \alpha'_s)^{-1}$ in terms of the Regge
slope, $\alpha'_s = l_s^2$. Another common convention defines $g^2_{YM} =
2\pi g_s$.)  This highly symmetric case is amenable to many analytic tools
which have greatly elucidated String/Gauge duality. However pure $SU(N_c)$
Yang-Mills for quarkless QCD has much less symmetry. There is no super
symmetry and conformal symmetry is
broken by quantum effects giving rise to ``dimensional transmutation'',
asymptotic freedom and confinement. Backgrounds that break conformal and some
or all SUSY symmetries differ in the precise details of the ``warping'' of
the 5th dimension.

Yang-Mills operators correspond to boundary conditions on the string (or
super gravity) Green's function at $r \rightarrow \infty$. It is sometimes
more convenient to invert the radial co-ordinate by introducing $z = R^2/r$,
placing this boundary at $z=0$,
\be
ds^2 = \frac{R^2}{ z^2}(\eta_{\mu\nu} dX^\mu dX^\nu +  dz^2) + R^2
d^2\Omega_5 \; .
\ee
Surprisingly, the conformal example exhibits stringy physics in spite of its
lack of confinement (and the stringy idea of narrow electric flux tubes
in the Yang-Mills description). Due to conformal symmetry the potential
between two heavy ``quark'' sources at separation L is purely  Coulombic, 
\be 
E_0(L) = -\frac{ (2 \pi)^2 }{\Gamma^4 (1/4)} \frac{\sqrt{ 2 g^2 N_c}}{L} \;.
\ee 
That is, the Wilson loop has perimeter rather than area law
behavior even as L goes to infinity.  At short distances this potential
is not a bad description of QCD up to the asymptotically free logarithmic
corrections and there is a discrete set of excited modes above the ground
states as described in Sec.~\ref{sec:stretched}. 

A useful prototype for a class of confining deformations can be expressed as
\be \label{eq:conf_metric}
ds^2 = \frac{R^2}{z^2} [\;\eta_{\mu\nu} dX^\mu dX^\nu + G^2(z/z_{max})dz^2
\; ]  +
\mbox{compact dimension}
\ee
which approaches $AdS^5$ in the UV, ($G^2(z/z_{max} \rightarrow 0) = 1$), but
is strongly modified  in the IR with a cutoff at $r_{\min} = R^2/z_{max}$.
The new mass scale, $\Lambda_{IR} \equiv r_{min}/R^2$, is proportional to
the mass gap or lowest glueball mass in the spectrum and plays the role of
$\Lambda_{qcd}$.  The simplest phenomenological model, exploited extensively
in the work of Polchinski and Strassler~\cite{PolStrass2}, uses the pure
$AdS^5$ metric for $r \ge r_{min}$ with a ``square-well'' cutoff at $r =
r_{min}$ (Dirichlet condition on super gravity fields),
\be
G_{cutoff}(z/z_{max}) = \theta(1- z/z_{max}) \; .
\ee
Here we use a softer IR cutoff inspired by an  $AdS^{\gamma + 1}$
Euclidean black hole confining background metric,
\be
G_{blackhole}(z/z_{max}) = \frac{1}{\sqrt{1 -  (z/z_{max})^\gamma}} \; .
\ee
\label{eq:BHmatrix}
A major consequence of this IR cutoff besides obviously introducing a
scale that breaks conformal symmetry is confinement.

\section{Qualitative Results for Confining Backgrounds}

The early calculations in a confining background included the glueball
spectrum~\cite{csaki, jev, bmt1,BMT} (at strong coupling) and a heuristic
picture of wide angle scattering~\cite{PolStrass1,BT}. By comparing these
calculations with lattice spectra for glueballs on the one hand and 
experimental data for hard scattering on the other (interpreted by parton
models at weak coupling), one can begin to understand the qualitative physics
of QCD like string models in deformed AdS space and their limitations. For
instance, the glueball wave functions are concentrated near $r = r_{min}$,
reflecting IR physics, whereas hard scattering ($s \sim -t$) is dominated
by $r^2 \ge r^2_{scatt} \sim \alpha'_s R^2 s$. Together they probe two
opposite limits of the 5th dimension~\cite{BPST}.

\subsection{Glueball Spectra}

\begin{figure}[ht]
\raisebox{0.42cm}{\includegraphics[width=2.75in,height=4.0in]{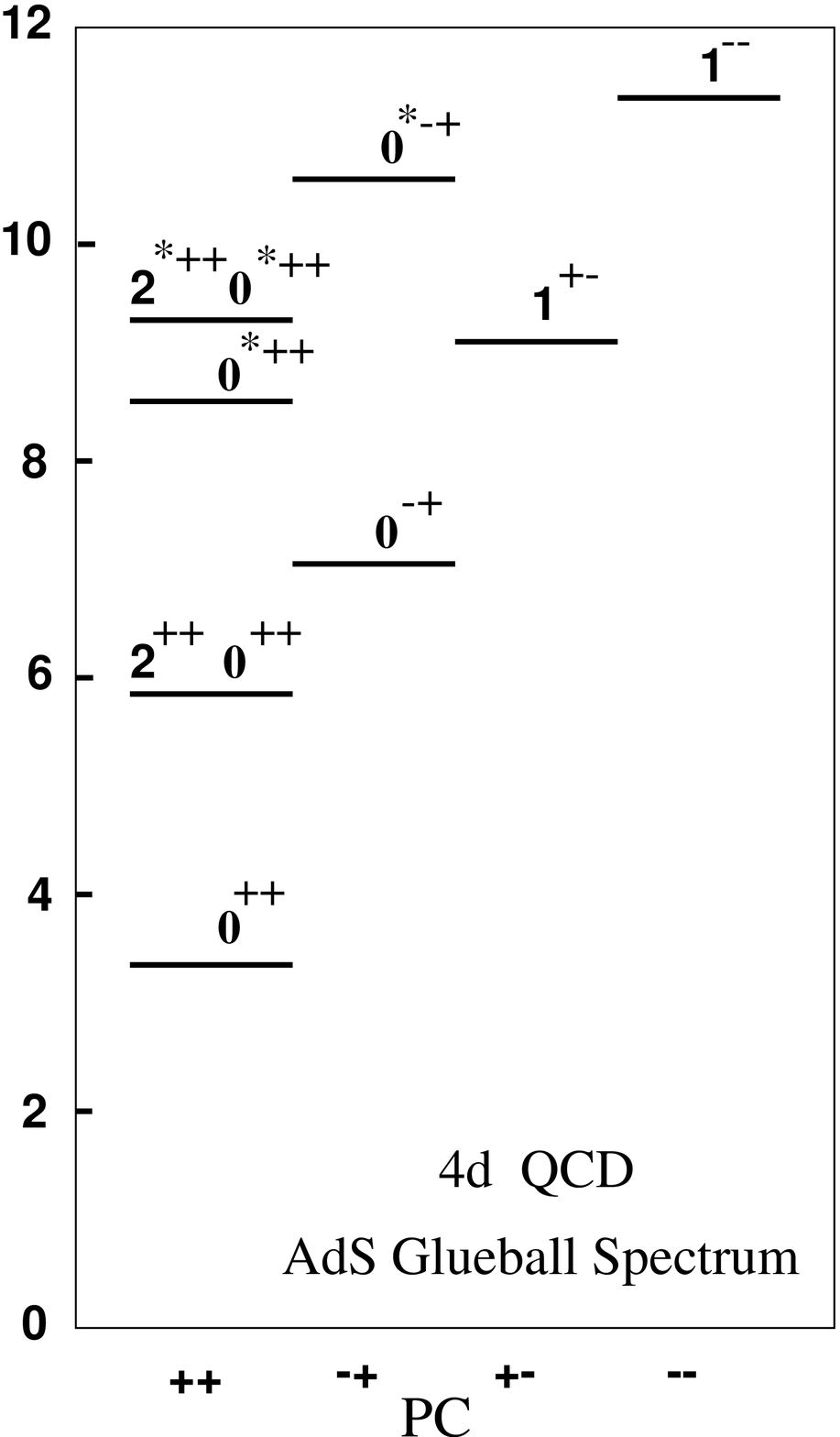}}
\includegraphics[width=3.25in,height=4.20in]{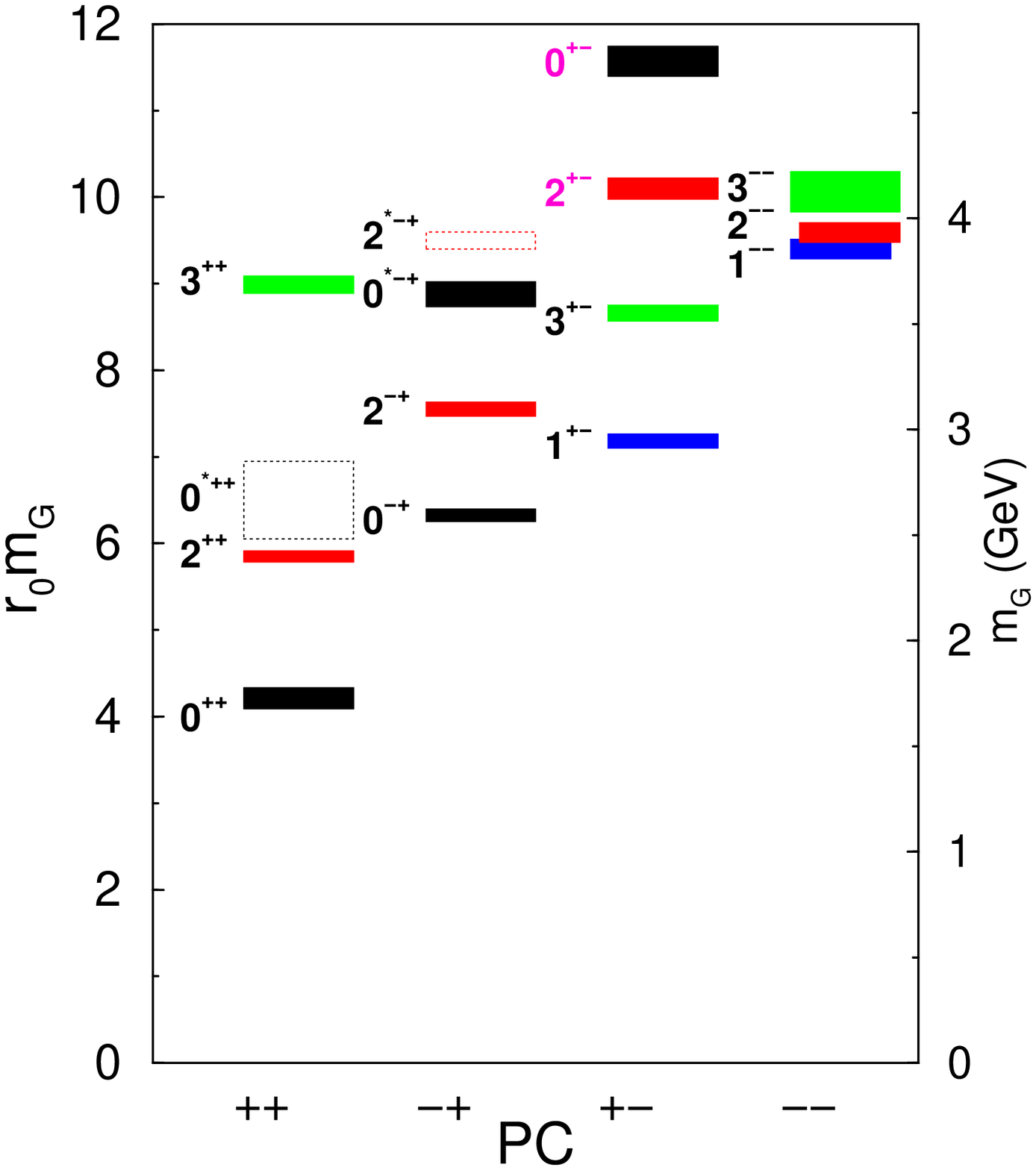}
        \caption{The AdS glueball spectrum for $QCD_4$ in strong
        coupling (left) compared with the lattice spectrum~\cite{MP} for pure
        SU(3) QCD (right). The AdS cutoff scale is adjusted to set the
        lowest $2^{++}$ tensor state to the lattice results in units of
        the hadronic scale $1/r_0 = 410$ Mev.}  
\label{fig:comparison}
\end{figure}

The computation of the glueball spectrum in strong coupling for the $AdS^7$
black hole (i.e., Eq.~\ref{eq:BHmatrix} with $\gamma =
6$,~\cite{BMT}) provided the first detailed application of String/Gauge
duality for confining theories.  From the effective Born-Infeld action on
the brane,
\be
S=\int d^5x
\det[G_{\mu\nu}+e^{-\phi/2}(B_{\mu\nu}+F_{\mu\nu})]+\int d^4x (C_1
F\wedge F+ C_3\wedge F+ C_5) \; ,
\ee
one can read off the quantum numbers in the dual Yang-Mills theory.  The
entire spectrum for all states in the QCD super selection sector were then
compared with lattice data~\cite{MP} for SU(3). In spite of the crude
approximations of strong coupling, the results are rather encouraging (see
Fig.~\ref{fig:comparison}). All the states are in the correct relative order
and the missing states at higher J are a direct consequence of strong
coupling which pushes the string tension to infinity.
It appears plausible that the $AdS^7$ black hole phase at strong
coupling is rather smoothly connected to the weak coupling (confined)
fixed point of QCD.  However it must be emphasized that there is no
general understanding of how the metric can be deformed so that all
the unwanted charged Kaluza-Klein states in the extra compact
directions decouple. So far all attempts to find better background solutions
to supergravity as a starting point for QCD have failed in this
regard.

\subsection{Near-Forward Scattering and Regge Behavior}

An important paper by Polchinski and Strassler~\cite{PolStrass1} began to
explain how hard scattering can be accommodated in the string description in
spite of the exponentially soft behavior for all string scattering amplitudes
in flat space. By assuming a single local scattering in a cutoff $AdS^5$
background, the glueball elastic scattering amplitude, $T(s,t)$, takes the
form $T(s,t) = \int_{r_h}^{\infty} dr \;{\cal K}(r) \> A(\hat s, \hat t, r)$,
where $A$ is a local 4-point amplitude for string scattering at $r$ and
${\cal K}(r) $ accounts for the phase space and glueball wave functions. In
the local frame for the string scattering amplitude, the Mandelstam
invariants are $\hat s = (R/r)^2 s$ and $\hat t = (R/r)^2 t$ , due to the
red-shift in r.

In the Regge limit the amplitude becomes
\be 
T(s,t) = \int_{r_{h}}^{\infty} dr\; {\cal K}(r) \> \beta( \hat t)
(\alpha'_s \hat s)^{\alpha_0 + {\half \alpha'_s}\> \hat t }\; .
\label{eq:regge}
\ee
As further explained by Brower and Tan~\cite{BT}, for small $t\simeq 0$, this
corresponds to an exchange of a BFKL-like Pomeron, with a small effective
Regge slope,
\be \alpha'_{BFKL}(0)\sim
\half (r_{min}/r_h)^2\alpha'_{qcd}<< \alpha'_{qcd}.  
\ee
Such an exchange naturally leads to an elastic diffraction peak with
little shrinkage~\cite{tan}. In the coordinate space, one finds, for a hard
process, the transverse size is given by
\be 
<{\vec X}^2> \sim \half (r_{min}/r_h)^2 \alpha'_{qcd} \log s +{\rm constant} \; . 
\label{eq:xsq} 
\ee 
If the cutoff, $r_h$, which characterizes a hard process, increases
mildly with $s$ ( e.g. $r_h^2\sim \log s$), there will be no transverse
spread.  

They also note that glueball wave functions are large in the IR so that there
is a strong scattering amplitude for $r \simeq r_{min}$,
\be T(s,t)\sim A( s, t, r_{min}) \sim (\alpha'_{qcd} s ) ^ {\alpha_{P}(0) +
  \half \alpha'_{qcd} t } \; ,  
\ee 
approximating the soft Regge pole~\cite{BT}.  The QCD Regge slope  to first
non-trivial order in strong coupling is
\be
\alpha'_{qcd} =  \frac{R^2}{r^2_{min}} \alpha'_s =
\frac{1}{\Lambda^2_{IR}}\frac{ \alpha'_s}{ R^2 } =  \frac{1}{ \Lambda_{IR}^2
  \sqrt{g^2 N_c}}
\ee
consistent with the surface tension of the open string, $T_{qcd} = (2 \pi
\alpha'_{qcd})^{-1}$, computed by the area law for a Wilson
loop~\cite{maldacena,tension,several}.  The linear extrapolation from the
glueball gives an estimate for the intercept of
\be
\alpha_P(0)= 2- \half \alpha'_{qcd} M^2_{GB} = 2 -  
\frac{2 M^2_{GB}}{\Lambda^2_{IR} \sqrt{g^2N_c}} \; ,
\ee
where $M^2_{GB}$ is the mass of the tensor $2^{++}$ glueball.  Of course, the
true Regge pole, that gives the zero-width physical states at $N_c =\infty$
string theory, must give a pure Regge power at fixed $t$ with exact
factorization in the t-channel~\cite{tan} in contradiction with the
approximate amplitude in Eq.~\ref{eq:regge} above. This cannot be achieved in
the local string scattering approximation at fixed r. Departures from this
local approximation are also necessary to really probe  partonic
features in deep inelastic scattering~\cite{PolStrass2}, since constituent
pieces of the string need to be isolated in the UV region. The consequences
for diffractive scattering and the BFKL Pomeron of going beyond the local
approximation is under active investigation~\cite{BPST}.

\section{Stretched String Spectra}
\label{sec:stretched}

\begin{figure}[ht]
\begin{center}
\includegraphics[width=4.0in,height=4.0in]{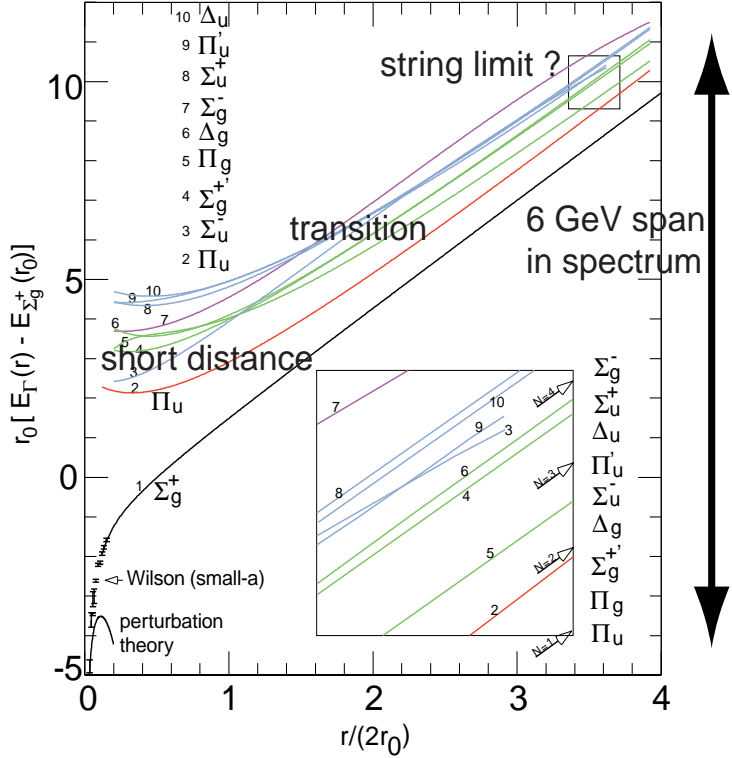}
\end{center}
\caption{Lattice string spectrum for SU(3) Yang-Mills theory
determined by Juge, Kuti and Moringstar~\cite{kuti}}
\label{fig:stringspec}
\end{figure}

An even more direct probe of the 5-th dimension is provided by the spectrum
of the string stretched between infinitely heavy sources (see
Fig~\ref{fig:stringspec}).  From the $AdS/CFT$ viewpoint, starting with the
ends of the string separated by a small distance L, we are probing the 
short distance Coulomb regime. Then as we increase L, the minimal surface
moves into the interior probing more and more IR physics. Finally at very
large L we see only the lowest mass transverse ``Goldstone modes'' of the
string leading to the universal spectrum of L\"uscher,
\be
E_n =  T_0 L + \frac{\pi}{L}[a^{\perp \dagger}_n a^\perp_n  - \frac{(D-2)}{24} ] + \cdots
\ee
Indeed at large separation L the lattice data for the stretched string
spectrum appears to be approaching this form with 2 transverse
oscillators, D-2 = 2. (See Fig.~\ref{fig:stringspec}). 
Using a very cleaver  lattice simulation method, L\"uscher and
Weisz~\cite{Luscher-Weisz} were apparently able to determine the one loop
contribution ( i.e.the so called  ``L\"uscher term'') to the ground state,
\be
 E^{1 loop}_0(L)  = - \frac{\pi}{12}(1 + 0.12fm/L) \; ,
\ee
for L in the range 0.5 to 1.0fm .  The agreement with theory for $D-2 = 2$ is
remarkable (if not paradoxical) in view of large
non-universal features in the spectrum for $L \simeq 0.5-1.0fm$
clearly visible in the lattice data (Fig~\ref{fig:stringspec}).

It is evident from the lattice data in Fig.~\ref{fig:stringspec} that a major
challenge for the AdS/CFT approach to the QCD string is to understand the
highly non-trivial interpolation between IR (large L) and UV (small L)
physics of the string.  As a first attempt, we are quantizing the string with
Dirichlet boundary conditions in the confining $AdS^5$ blackhole metric of
Eq.~\ref{eq:conf_metric} above. First we solve (numerically) for the minimal
surface getting a classical potential energy, $ E^{class}_0(L)$ for the
ground state of the stretched string.  This function obeys the following
limits,
\be
 E^{class}_0(L\rightarrow \infty)  \simeq
 \frac{r^2_{min}}{2 \pi \alpha'_s R^2} \; L + O(Le^{- c L}) \quad \mbox{and}
 \quad E_0^{class}(L\rightarrow 0) \simeq - \frac{ \pi^2 \sqrt{2}}{
  \alpha'_s\Gamma(1/4)^4 } \; \frac{R^2}{L} 
\ee
The exact function,$ E^{class}_0(L)$, fits almost perfectly the lattice data
for all L, after adjusting the mass scale $R^2/r_{min} = \Lambda_{IR}$ and
the Regge slope $\alpha'_{qcd} = R^2 \alpha'_s/r^2_{min}$.  This is
reassuring but also highlights the present weakness of our AdS model. Pure
Yang-Mills (quarkless QCD) predicts a definite number for the string tension,
$T_{qcd} = 1/(2 \pi \alpha'_{qcd})$ relative to the single Yang-Mills scale,
$\Lambda_{qcd}$.  However at present all AdS/CFT strong coupling models for
confining theory have an extra cutoff parameter that can be adjusted 
independent of the Regge slope (or QCD string tension).  This extra mass
scale is similar to strong coupling lattice QCD, except that on the lattice
the new scale is the UV cutoff of the lattice spacing whereas here it is a
phenomenological IR scale more similar to the MIT bag constant.

\begin{figure}[ht]
\begin{center}
\input{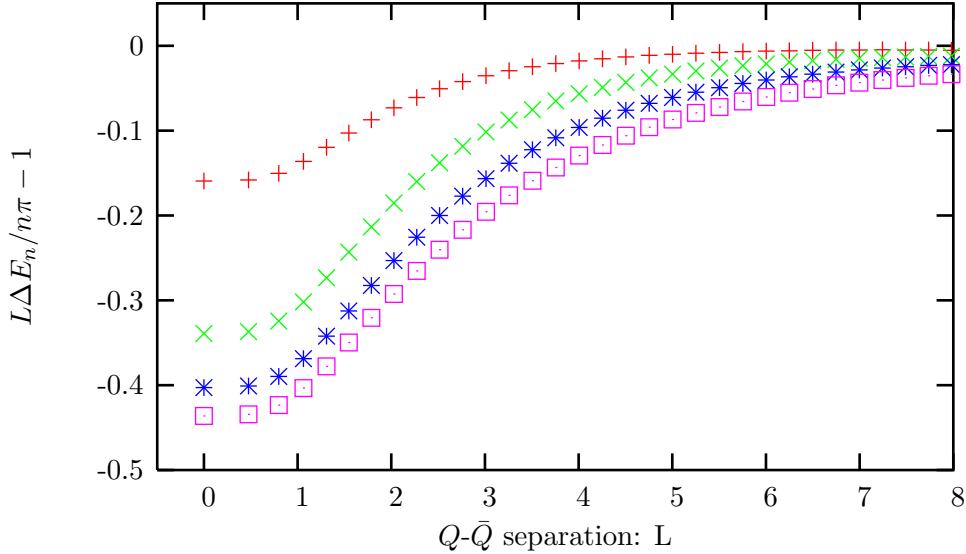}
\end{center}
        \caption{Deviations of transverse energy levels, $L \Delta E_n/(n
          \pi) - 1$, in confining $AdS^5$ black hole
          model for n = 1 ($+$), 2 ($\times$),  3 ($*$) and 4 ({\tt
           $\Box$}).}
\label{fig:stretched}
\end{figure}

Work is currently underway to compute the energy levels, $E_n = E^{class}_0 +
\Delta E_n$, above the ground state in our model AdS blackhole background.
In a temporal gauge $X^{(0)} = t$ with $\sigma = X^{(3)}$, the linearized or
semi-classical equation for the transverse fluctuations of a string stretched
is
\be
[\; \partial^2_t - v^2(\sigma) \partial^2_\sigma\; ]\; X^\perp(t,\sigma) = 0 \; ,
\ee
The local velocity, $v(\sigma) = z^2(\sigma)/z^2_c$, of propagation along the
string is bounded by the speed of light. At large L, the velocity approaches
a constant, exponentially close to the speed of light, except near the quarks
at $z = 0$. Thus, it is not surprising that at large L this
confining background reproduces flat string results, 
$$\Delta E_n = \frac{n \pi}{L}\; , $$
and in one loop, the universal L\"uscher term (or Casimire energy), $\Delta E^{1
  loop}_n = (D-2) \pi/24$. Moreover a numerical solution to our equation in
Fig.~\ref{fig:stretched} shows deviations from the large L asymptotic  value in
qualitative agreement with that measured in the Lattice simulations of Juge, Kuti
  and Morningstar~\cite{kuti}.
However there is another contribution to this deviation with the same sign
observed in the exact spectrum of the (flat space) Nambu-Goto
\be
E^{Nambu-Goto}_n = T_0 L \sqrt{1 + \frac{2\pi}{L}[a^{\perp \dagger}_n a^\perp_n  - \frac{(D-2)}{24} ]}
\ee 
when quantized in the lightcone gauge~\cite{arvis}. An effort is
underway to combine these two effects in a self-consistent lightcone
quantization of the string in a confining AdS space~\cite{brower-tan-thorn}.

As you approach zero separation $L \rightarrow 0$, the discrete spectrum of
the stretched string is determined by the pure $AdS^5$
metric~\cite{brower-tan-thorn}. The exact spectrum is know in closed form
through the sum rule, 
\be
z_c \Delta E_n  \sqrt{(z_c \Delta E_n)^4 - 1} \int^1_0 \frac{ds}{[1 + (z_c
  \Delta E_n )^2]
\sqrt{(1- s^2)}} = \frac{n \pi}{2} \; ,
\ee
give by Callan and Guijosa~\cite{callan} where $n=1,2,\cdots$ and $z_c = (2
\pi)^{-3/2}\Gamma(\quarter) L $. In the conformal limit the energies must of
course be proportional $1/L$.

In addition to transverse Goldstone modes, there are quantum modes for
fluctuations in the extra ``radial'' direction,
\be
[\; \partial^2_t - v^2(\sigma) \partial^2_\sigma\; ]\; \rho(t,\sigma) = M^2(\sigma)
\rho(t,\sigma) \; .
\ee
Unlike the transverse Goldstone modes $X^\perp$, now there is a
$\sigma$-dependent ``rest mass'',
\be
M^2(\sigma) = \frac{1}{z^2_c} [z \frac{ d}{dz}\frac{1}{G^2} 
              -  \frac{2 z^4}{z^4_c G^2}] \; ,
\ee
whose scale is set by the glueball mass: $M_{GB} \sim 1/z_c$.  These modes
correspond via the String/Gauge duality to longitudinal (breathing) modes for
a fat chromodynamic flux tube. It would be interesting to find these modes in
lattice gauge theory simulations and trace their dependence on L.  

Still our toy QCD string in an $AdS^5$ blackhole is at best just a first step
in understanding how a QCD string in warped space might behave.  Much work
remains even to identify the microscopic degrees of freedom of the QCD string
let alone to the discovery of an effective string action capable of
reproducing the lattice spectrum from long distance into the short distance
region governed by asymptotically free gauge theory at large $N_c$.  However
a reasonable near term goal is to find accurate interpolation formulae for
all L for the low energy spectrum of the stretched sting in a legitimate
confining super gravity background. On the basis of the comparison of this
spectrum with lattice data one might narrow the search for the QCD background
geometry itself.

Acknowledgment: We wish to thank Joe Polchinski, Matt Strassler and Charles
Thorn for very interesting discussions during our visit to the KITP Workshop
on {\it QCD and String Theory} where this draft was written. Also the work on
the string spectrum benefited substantially from our interactions with Oliver
Jahn and Carlos Nu\~nos at Massachusetts Institute of Technology.


\begin{thebibliography}{9}
\bibitem{BPST} Current work in 
collaboration with J. Polchinski and M. Strassler is
revealing a much clearer understanding of how to probe AdS in the Regge
scattering limit. Also see Ref. ~\cite {PolStrass2}  for the use of DIS as a probe.
\end{thebibliography}
\end{document}